\documentstyle[12pt,epsf,axodraw]{article}
\def\gsim{\mathrel{\rlap {\raise.5ex\hbox{$ > $}}
{\lower.5ex\hbox{$\sim$}}}}
\def\lsim{\mathrel{\rlap {\raise.5ex\hbox{$ < $}}
{\lower.5ex\hbox{$\sim$}}}}

\newcommand{\be}{\begin{equation}}
\newcommand{\ee}{\end{equation}}
\newcommand{\bea}{\begin{eqnarray}}
\newcommand{\nn}{\nonumber}
\newcommand{\eea}{\end{eqnarray}}

\def\gappeq{\mathrel{\rlap {\raise.5ex\hbox{$>$}}
{\lower.5ex\hbox{$\sim$}}}}
 
\def\lappeq{\mathrel{\rlap{\raise.5ex\hbox{$<$}}
{\lower.5ex\hbox{$\sim$}}}}
 
\begin{document}
 
\begin{titlepage}
\begin{flushright}
OUTP--98--25P \\
hep-th/9803189 \\
\end{flushright}

\begin{centering}
\vspace{.1in}
{\large {\bf World-Sheet Defects, Strings, and Quark Confinement$^{*}$}} \\
\vspace{.2in}

{\bf N.E. Mavromatos$^{a}$} \\

Department of Physics
(Theoretical Physics), University of Oxford, 1 Keble Road,
Oxford OX1 3NP, U.K.  \\

\vspace{.5in}
 
{\bf Abstract} \\
\vspace{.1in}
\end{centering}
{\small In this talk I give a preliminary account of original  results,
obtained in collaboration with John Ellis. Details and further elaboration  
will be presented in a forthcoming publication~\cite{em}. 
We present a proposal for a non-critical (Liouville) string approach to 
confinement of four-dimensional (non-abelian) gauge theories, based on 
recent developments on the subject by Witten and Maldacena. 
We discuss the effects of vortices and monopoles 
on the open  world-sheets whose boundaries are   
Wilson loops of the target-space (non Abelian) Gauge theory.
By appropriately employing `D-particles', associated with the 
target-space embedding of such defects, we argue that the appearance of 
five-dimensional Anti-De-Sitter (AdS) space times is quite natural, as a 
result of Liouville dressing, required for a consistent description of 
the D-particle quantum fluctuations (`recoil'). 
The D-particle `recoil' may be considered as a dual description 
of the `distortion of  space-time' caused by 
the propagation of the heavy test particle 
along the Wilson loop of the target-space gauge theory. 
We isolate the world-sheet defect contributions to the 
Wilson loop by constructing an appropriate observable; this observable is 
the same as the second observable 
in the supersymmetric U(1) theory of Awada and Mansouri, 
but in our construction supersymmetry appears not necessary.
When vortex condensation occurs, we argue in favour of a (low-temperature) 
confining phase, in the sense of an area law, for a large-$N_c$ 
(conformal) gauge theory at finite temperatures. A connection  
of the Berezinski-Kosterlitz-Thouless (BKT) transitions on the world-sheet 
with the critical temperatures in the thermodynamics of Black Holes 
in the five-dimensional AdS space  is made. The upper critical 
temperature in that case is identified with the monopole 
condensation BKT temperature, which is 
argued to be responsible for setting the scale for new Physics.}

\vspace{1.in}
\begin{flushleft}
$^{a}$ P.P.A.R.C. Advanced Fellow. \\
$^{*}$Based on talk to 
be presented by the author at the International Workshop of 
the `Hellenic Society 
for the Study of High Energy Physics',  Democritos NRC, Athens, Greece, 
April 8-11 1998.

\end{flushleft}

\end{titlepage} 

\newpage

\section{Introduction}

The recent exciting developments in the understanding of
non-perturbative effects in the theory formerly known 
as strings~\cite{dbranes}
have led to tantalizing glimpses of a broader framework
for the Theory of Everything (TOE),
referred to variously as $M$ or $F$ theory. 

Moreover, quite recently, some of these developments, 
have been argued to play an important r\^ole 
in a non-perturbative description 
of the long-distance Physics 
of certain large $N$ Gauge theories, and in 
particular in their confining aspects~\cite{witten}.
In the center of these developments lies the conjecture by
Maldacena~\cite{malda},
that certain large $N$ conformal field theories may be 
understood as being related to Supergravity 
or String theory in the bulk of certain Anti-de-Sitter ($AdS$) space times,
whose boundary is a conformal Minkowski space-time, on which 
the large $N$ conformal field theory lives. 

Formally, the conjecture may be formulated as follows~\cite{witten}.
Let $Z_S(\phi _0)$ be ye supergravity of (super)string action 
in the bulk of a (d+1)-dimensional 
$AdS_{d+1}$, computed with the boundary condition that 
at the boundary of $AdS_{d+1}$, $M_d$ (a conformal Minkowski
space time), the field $\phi$ approaches $\phi _0$. 
The ansatz for the Conformal-Field-theory CFT/AdS correspondence,
proposed in \cite{malda}, and elaborated in \cite{witten}, 
can then be stated as:
\be
\langle {\rm exp}\int _{M_d} \phi_0 {\cal O} \rangle =Z_S(\phi _0)
\label{conjecture}
\ee
where ${\cal O}$ are operators in the CFT. The important 
property of $AdS$ is its `holographic' nature, in the sense
of uniqueness theorems~\cite{lee,witten}, that 
specify the bulk behaviour of the 
classical field (or string) theory of $\phi$ in terms of 
the boundary value $\phi_0$.

The conjecture (\ref{conjecture}) allows for 
a (non-perturbative) computation 
of correlation functions of certain supersymmetric conformal field theories,
notably some conformal supersymmetric gauge theories in their 
strong coupling (confining) regime~\cite{witten}. 
In view of (\ref{conjecture}) and the associated holographic nature of $AdS$,
this means that quantum information about the confining physics of 
a non-Abelian Gauge theory is thus encoded in classical geometries.  
It should be noticed that in order for the supergravity solutions 
used in \cite{malda,witten} to be trusted one must work 
in a large $N_c$ limit of a strongly coupled 
supersymmetric and conformal $U(N_c)$ gauge theory.
The appropriate limits are taken in such a way that 
$g_{YM}^2 N_c $ is fixed but {\it large}, as $N_c \rightarrow \infty$,
with $g_{YM}^2 \sim g_S$, with $g_S$ the string theory coupling.  
The supersymmetry in the above approach is needed
because the entire approach is based on critical-dimension 
`super-string' theory, 
in the sense of tensoring the four-dimensional 
space-time manifold 
with appropriate compact manifolds, so as to 
ensure a critical total central charge.

It is the purpose of this talk to 
report on an attempt 
to extend these developments
to a non-critical-dimension string theory, which might open the way 
for an extension of the above ideas to non-supersymmetric theories. 
This talk is based on original research done in collaboration with 
John Ellis, and a detailed account of it
will appear in a 
forthcoming
publication~\cite{em}. 
We make a proposal concerning a Liouville string description of 
(non-supersymmetric) four-dimensional Gauge theories, which are 
conformal at zero temperatures. 
One of the most important points to make is that in such attempts
one faces the necessity to include in the world-sheet of the 
string, whose boundary is the Wilson loop of the pertinent Gauge theory, 
non-trivial defects (monopoles or vortices).
The world-sheet thermodynamical properties of such defects 
will be argued to dictate the various phases of 
the finite-temperature target-space 
Gauge theory, through a non-trivial correspondence
with the embedding $AdS$ space times, arising as a result of appropriate 
Liouville dressing. 

The layout of the talk is as follows: In section 2, we review
aspects of our previous analysis of Liouville 
representation of 
world-sheet defects.
In section 3 we 
discuss the connection of AdS space times and D-brane recoil Physics,
as well as how (conformal) QCD, viewed 
 as a non-critical string theory, could fit into this 
picture. In section 4 we 
present the Awada-Mansouri approach~\cite{awada} 
to the connection between 
Gauge theories and scale invariant (super)strings in target space,
which we shall make use of in our attempt to discuss 
a dynamical scenario for the appearance of the string tension 
in (conformal invariant) string theories. 
In section 5 we discuss the r\^ole of world-sheet defects
on the (long-distance) physics of quark confinement. 
We argue, in particular, that the 
presence of such defects on the world sheet 
makes the role of supersymmetry in the approach of \cite{awada} 
redundant~\footnote{In fact 
finite temperature field theories which are 
discussed in the recent literature~\cite{malda,witten}
have their target space supersymmetry (and conformal symmetry) broken.}. 
We also give arguments 
in support of the r\^ole of world-sheet vortices and monopoles 
on inducing the confining aspects
of large-$n$ $U(n)$ gauge theory, in the sense of  
an area law, and 
a non-perturbative computation of the string tension. 
We thus relate the various Berezinski-Kosterlitz-Thouless (BKT)
transitions on the world-sheet to the thermodynamics of AdS Black Holes,
which has been argued~\cite{witten} to be relevant for the various phases
of the Gauge theory. 
We also present some remarks clarifying
the r\^ole of the Abelian projection to confinement in pure
glue Yang-Mills theories.  
Conclusions and outlook are presented in section 6.

\section{World-Sheet Defects in (Liouville) Superstring Theories}

\subsection{General Remarks}

In this section we shall review some basic features of world-sheet
defects, which we shall use in our approach.
A vortex defect on the world 
sheet~\cite{sathiap,emnmonop} is 
obtained as the solution $X_v$ of the equation
\begin{equation}
\partial_z {\bar \partial}_z X_v = {i \pi q_v \over 2} [ \delta(z - z_1) -
\delta(z - z_2)]
\label{defect}
\end{equation}
where $q_v$ is the vortex charge and $z_{1,2}$ are the locations of
a vortex and antivortex, respectively, which we may map to the
origin and the point at infinity. The corresponding
solution to (\ref{defect}) is
\begin{equation}
X_v = q_v {\rm Im~ln} z
\label{solution}
\end{equation}
and we see that the vortex charge $q_v$ must be integer. 

There are related `spike' configurations which are solutions of the
equation
\begin{equation}
\partial_z {\bar \partial}_z X_m = - {\pi q_m \over 2} [ \delta(z - z_1) -
\delta(z - z_2)]
\label{spike}
\end{equation}
given by
\begin{equation}
X_m = q_m {\rm Re~ln} z
\label{solution2}
\end{equation}
It is easy to see~\cite{ovrut} 
that,
in the presence of {\it both} types of defects, 
single-valuedness of the partition function imposes
the following quantization condition:
\begin{equation}
2 \pi \beta q_v q_m = {\rm integer}
\label{qcond}
\end{equation}
at finite temperature $T \equiv \beta^{-1} \ne 0$.

There is a vortex-monopole duality, which is manifested as
the invariance of the defect partition function under~\cite{cardy}: 
\be
   \pi \beta \leftarrow\rightarrow \frac{1}{4\pi\beta}~~; \qquad 
q_v \leftarrow\rightarrow q_m 
\label{exchange}
\ee

Consider now an Abelian gauge theory 
description of the defect dynamics, such that the 
world-sheet vector field is the pullback  
of
some background 
gauge field $A_M(X)$, taken to be 
Abelian for definiteness, over which the string propagates.
The `pullback' of 
the gauge field on the world-sheet can be defined as:
\be
  A_\alpha (z, {\overline z})=\partial_\alpha X^M A_M (X) 
\label{pullback}
\ee 
where Greek indices are world-sheet variables, $\alpha =1,2$, and 
upper case Latin indices $M=1, \dots D$ are target-space indices. 
The world-sheet `magnetic field' corresponding to 
(\ref{pullback}) reads: 
\be
{\cal B} = \epsilon^{\alpha\beta}\partial_{\alpha}A_\beta =
\epsilon^{\alpha\beta}\partial_{\alpha}\partial_\beta X^M A(X)_M 
+ \epsilon^{\alpha\beta}\partial_{\alpha}X^M \partial_\beta X^N 
\partial_N A_M (X) 
\label{pullback2}
\ee
Consider now the case where 
the world-sheet surface has the topology 
of a disc, whose boundary $C$ will be identified in our work with a 
Wilson loop for the field, $A_M$, in target (embedding) space:
\be
     W(C)=e^{ie\int _C A.dl}
\label{wl}
\ee
In such a case the world-sheet flux of 
${\cal B}$ through the world-sheet $\Sigma (C)$, can be 
identified with the flux of the target-space gauge field
$A_M$, in the embedding space. 
By the Stokes theorem on the world-sheet, we then see that 
in the flux $\int _{\Sigma (C)} {\cal B}.dS= \int _C A.dl$,
the 
contribution from the first term on the r.h.s. of (\ref{pullback2})
can be attributed to vortices $X^v$ on the world-sheet, 
whilst the second contribution to monopoles, $X^m$. 
This is in accordance with the general Hodge 
decomposition 
of an arbitrary gauge field $A_\alpha$ on $\Sigma (C)$.
It is the second term which corresponds to target-space monopoles
for the field $A$, since 
it is gauge invariant in target space,
and enters 
the computation 
of space-time fluxes~\footnote{As a side remark, 
it is interesting to note that 
the quantity 
$\sigma^{MN} \equiv \epsilon^{\alpha\beta}\partial_\alpha X^M 
\partial_\beta X^N $
enters the Schild action~\cite{schild}, recently 
claimed~\cite{yoneya} to be associated with a matrix-model representation 
of D-brane dynamics~\cite{dbranes}. As we shall discuss, in our case, 
the relevant quantity for the confining aspects of gauge theory 
is $\sigma ^M \equiv \epsilon^{\alpha\beta}\partial_\alpha\partial_\beta X^M$,
which is also non-vanishing for a vortex (angular field), or at the monopole core.}. 

In this way, the presence of a magnetic monopole in the embedding space
may be pull-backed as a monopole (or vortex) defect 
of the (world-sheet) 
surface $\Sigma (C)$. This will always 
by understood in our discussion in this work.
We also note that in the context of two-dimensional 
target-space embeddings, 
world-sheet monopole defects can be 
mapped~\cite{emnmonop,emn} 
to topologically non-trivial two-dimensional 
target space times
with Schwarzschild 
black holes\cite{wittenbh}.
In the higher-dimensional case, 
discussed in this work, we shall find an analogous phenomenon:
the physics of world-sheet monopoles will be associated 
with the physics of Black Holes in (five-dimensional)  
Anti-de-Sitter (AdS) space times.

\subsection{Vortices and Monopoles in Liouville Strings}

The analysis of \cite{ovrut}, which we have adopted and extended
in our approach~\cite{emn}, 
indicated that it is possible to interpret
the Liouville theory~\cite{ddk}
in the {\it dangerous region} of the central charge 
$1< {\cal D} <9~(\rm or ~25)$ in terms of 
defect configurations
in a two-dimensional world-sheet theory with finite
(inverse) temperature, given by the appropriate 
`matter' central charge deficit~\footnote{However, one can also 
work in the `stringy' 
region for the central charge ${\cal D} > 9$, since the analysis of 
\cite{ovrut} applies in that case as well.}.  
For our purposes, of representing QCD 
(in some limits) as a non-critical string theory, we shall review
only the supersymmetric case. 

In our non-critical string 
interpretation, the `matter' central charge, ${\cal D}$, 
in general differs from the number of 
space-time dimensions~\cite{aben}, $D$, by the matter screening charge,
or, in general, by the `dilaton' fields:
\be  
      {\cal D} = D + 12 Q^2 
\label{cc}
\ee
In our case 
we assume $D=4$ (the target-space dimensionality 
of ordinary gauge theories).  
In such a case, one restores criticality by allowing a Liouville 
central charge deficit, $c_L$, such that 
$c_{total} = {\cal D} + c_L = 10$. 
The Liouville mode 
is `time-like'~\cite{aben}, if ${\cal D} > 9$ and space-like 
if ${\cal D} < 9$. The Liouville mode decouples at ${\cal D}=10$.

The supersymmetrization of the
world-sheet defects may be represented using a
sine-Gordon theory with local $n=1$ supersymmetry, which has
the following monopole deformation operator~\cite{ovrut};
\begin{equation}
V_m = {\bar \psi} \psi : {\rm cos} [{e \over \beta^{1/2}_{n=1}}
(\phi(z) - \phi({\bar z}))]:
\label{susymonopole}
\end{equation}
where the $\psi, {\bar \psi}$ are world-sheet fermions with
conformal dimensions $(1/2,0), (0,1/2)$ respectively, and
$\phi$ is a Liouville field. The effective temperature
$1 / \beta_{n=1}$ is related to the matter central charge by
\begin{equation}
\beta_{n=1} = { 2 \over \pi |{\cal D}-9|}
\label{susybeta}
\end{equation}

The corresponding vortex
deformation operator is
\begin{equation}
V_v = {\bar \psi} \psi : {\rm cos} [ 2 \pi q \beta_{n=1}^{1/2}
(\phi(z) + \phi({\bar z}))]:
\label{susyvortex}
\end{equation}
where $q$ is the vortex charge. The duality (\ref{exchange}) is 
valid also in this approach. 

Including the conformal dimensions
of the fermion fields, we find that the conformal dimensions of the
vortex and monopole operators are
\begin{equation}
\Delta_v = {1 \over 2} + {1 \over 2} \pi \beta_{n=1} q^2 = {1 \over 2} +
{q^2 \over |{\cal D}-9|}, \;\; \\
\Delta_m = {1 \over 2} + {1 \over 8 \pi \beta_{n=1}} e^2 = {1 \over 2} +
{e^2 |{\cal D}-9| \over 16}
\label{dimensions}
\end{equation}
respectively. 

Let us first consider the case ${\cal D} > 9$. 
We see that the supersymmetric vortex deformation 
with minimal charge $|q| = 1$ is marginal when ${\cal D} = 11$. 
Below this
dimension, the vortex deformation is irrelevant, and this is 
the case which shall interest us in the context of the 
present work. 
The simplest case is to consider minimum charge 
defects $|q|=1$, 
and study the various phases of such a world-sheet deformed theory,
which will be characterized by {\it different} values of the matter central charge
${\cal D}$.
The quantization
condition 
(\ref{qcond}) 
tells us that the allowed charge for a monopole defect, 
in the presence of a $|q|=1$ vortex, 
is $|e| = \frac{|m|}{4}({\cal D} - 9)$, $m \in Z$. 
In this case, for $m=1$
we find {\it three} regions~\cite{ovrut,emnmonop,emndbmonop}~\footnote{
It is understood that, in the general case,
the critical values depend on the charges $q,e$.}:
\bea
&~&  n=1~{\rm world-sheet~supersymmetry}: \nn \\ 
&~& 9 < {\cal D} < 11 ~~~~~{\rm monopole~vacuum~unstable,~vortices~bound} 
\nn \\
&~& 11 < {\cal D} < 14.04~~~~~~{\it both}~{\rm monopole~and~vortex~vacua~unstable}
\nn \\
&~& 14.04 < {\cal D} ~~~~{\rm ~monopoles~bound,~vortex~vacuum~unstable}
\label{regions}
\eea
The ${\cal D} >9$ case includes critical string theories, as well as 
their generalizations to M and F theory~\cite{emndbmonop}.

We next consider the case ${\cal D} < 9 $, which is most relevant for 
our low-energy description of gauge theories, as we shall discuss later on.
A similar analysis reveals the following phase diagram: 
\bea
&~&  n=1~{\rm world-sheet~supersymmetry}: \nn \\ 
 &~&  {\cal D} < 3.96 ~~~~~~~{\rm 
monopoles~bound,~vortex~vacuum~unstable} \nn \\
&~& 3.96 < {\cal D} < 7 ~~~~~{\it both}~{\rm monopole~and~vortex~vacua~unstable} \nn \\
&~&  7 < {\cal D} < 9 ~~~~~~~{\rm monopole~vacuum~unstable,~vortices~~bound}
\label{regions2}
\eea
Thus, in the ${\cal D} < 9$ case 
the monopole BKT transition occurs at lower values of ${\cal D}$ 
than the corresponding transition for vortices, which is the opposite 
of the situation encountered in the ${\cal D} > 9$ case.

However, from the point of view of BKT temperatures, in {\it both} cases
the temperatures (\ref{susybeta}) for the 
vortex condensation is lower than that of monopoles:
\bea
&~&  n=1~{\rm world-sheet~supersymmetry}/{\rm BKT~Temperatures}: \nn \\ 
&~&  T < T_{vortex}~~~~~~~{\rm 
vortices~bound,~monopole~vacuum~unstable} \nn \\
&~& T_{vortex} < T < T_{monop} ~~~~~{\it both}~{\rm monopole~and~vortex~vacua~unstable} \nn \\
&~&  T_{monop} < T < \infty ~~~~~~~{\rm monopoles~bound,~vortex~vacuum~unstable}
\label{regions3}
\eea
where the temperatures are understood to be given in units of the 
corresponding string tension. 
Notice, that as a result of 
monopole-vortex duality there is no 
region where both of these defects are stable.
 
In standard string theory~\cite{sathiap}, 
compactified 
on a circle $S^1$, the inverse temperature is 
the square of the compactification 
radius $\beta =R^2$ (in units of $\alpha '=1$). 
Usually in such a case~\cite{sathiap}
one considers only one type of 
defects, since the other can always be derived by a $T$ duality 
transformation. In that case the vortex condensation BKT temperature
is identified with the Hagedorn temperature,
where the winding modes of the string become tachyonic.
However, in our approach we shall be more general and consider
both types of world-sheet defects. As 
we have just seen, this leads to richer phase diagrams. 
In that case, for compactified (Bosonic) strings we have
for the vortex condensation~\cite{ovrut}: 
\be
     \frac{1}{2}\pi R_H^2q^2=1 ~~~{\rm if~|q|=1} \rightarrow R_H=\sqrt{2/\pi}
\label{haged}
\ee
whilst the monopole condensation occurs at:
\be
     \frac{e^2}{8\pi R_M^2}=1 ~~~{\rm if~|e|=1} \rightarrow R_M=1/(2\sqrt{2\pi})
\label{monkt}
\ee
From the 
above point of view, then, 
there appears to be an {\it intermediate temperature},
corresponding to the self-dual radius of compactification
\be
R_{1}=1/\sqrt{2\pi}
\label{sd}
\ee
Notice the following relation between $R_H$,$R_M$, and $R_1$:
\be
             R_1^4=(R_M R_H)^2
\label{connect}
\ee
We 
shall come 
back to a comparison of this phase diagram with the Thermodynamics 
of $AdS$ in section 5. 

The $n=2$ world-sheet supersymmetric case,
responsible for ${\cal N}=1$ target-space supersymmetry, 
is less interesting, in the sense that 
the vortex or  monopole vacua 
are stable, in the region $ {\cal D} > 1$, ${\cal D} < 1$, 
while the point ${\cal D} =1 $ appears as degenerate. 
So,
this theory has no such phase transitions. 
In \cite{emndbmonop} we have discussed ways of breaking
the $n=2$ supersymmetry down to $n=1$; 
from the target space-time point of view 
this implies breaking of the target ${\cal N}=1$ supersymmetry. 
In this work we shall be interested mostly in 
phase transitions at finite temperatures, where supersymmetry is broken.
From this point of view, the $n=1$ supersymmetric 
world-sheet case (or the corresponding bosonic, which, however, 
we do not discuss here) becomes
of interest. 

We have argued in \cite{dbrecoil}
that, in the context of string theory, 
the world-sheet 
 defects correspond to $D$(irichlet) branes~\cite{dbranes},
since correlators involving defects and closed-string
operators have cuts for generic values of $\Delta_{v,m}$.
These cause the theory to become effectively that of an open string.
One may then impose Dirichlet boundary conditions on the
boundaries of the effective world sheet, i.e., along the cuts,
obtaining solitonic $D$-brane configurations.
In this work we shall deal with QCD, and in particular 
we shall offer a scenario for a non-critical string representation.
In our approach Dirichlet particles, and in particular 
their `recoil'~\cite{kmw}, will play a crucial r\^ole. 
An association of QCD with non-critical
strings  has been advocated  
by Polyakov~\cite{polyakov}, but our approach will be different, 
since we shall use world-sheet defects, condensation of 
which will be argued to be responsible for the 
confining aspects of the (long-distance) physics of the 
non-Abelian gauge theory. 
In this respect, 
we shall associate the  regions (\ref{regions})  
to the confinement-deconfinement phase transition of QCD, 
following and extending the work of \cite{witten}, to 
connect the B-K-T world-sheet transitions 
to the critical temperatures 
of a gas of Black Holes in AdS space times~\cite{page}. 

To understand the connection with anti-de-Sitter space times 
let us first discuss how such space times can 
arise in our recoil approach to 
$D$ branes. The use of $D$-branes, in the dual string picture, 
may be seen as an alternative way of dealing with a gauge theory.
Indeed, it is widely accepted today~\cite{dbranes,lms} 
that $m$ parallel branes
have a gauge $U(m)$ symmetry, and their dynamics is equivalent
(in the sense of string dualities)
to the dynamics of certain supersymmetric Yang-Mills gauge theories
with group $U(m)$. Such issues are best understood in the 
regions where the number $m \rightarrow \infty$, so such an approach 
may be considered as a stringy picture of $U(N_c)$ gauge 
theories, with $N_c$ large.

From our point of view, the presence of a (dual) D-particle, or
a collection thereof, may also represent `back reaction' of the space-time 
due to the 
sudden movement of a heavy test particle (quark-antiquark pair) 
along the Wilson loop
in pure glue gauge theory; such a movement causes distortion 
of the surrounding space time, and the corresponding fluctuations 
can be described by the excitation of a D-particle `recoil' 
operator~\cite{kmw,dbrecoil,lizzi}, in a $\sigma$-model approach. 
Notice that the mass of such a particle is inversely proportional to 
the (dual) string coupling $g_s$, which therefore is weak. 
As we shall discuss below, the presence of the D-particle `recoil'
is crucial in yielding AdS space-times in a {\it dynamical way}. 

\section{Anti-de-Sitter Space Times from D-particle Recoil} 

\subsection{Recoil and Liouville Dressing: General Formalism}

In this section we would like to discuss how the anti-de-Sitter space time 
arises from our Liouville approach to D-brane recoil~\cite{dbrecoil}. 
The analysis was motivated originally by attempts, in collaboration
with J. Ellis and D.V. Nanopoulos~\cite{mth},  
to understand the structure of the elusive $M$-theory description of
`grand-unified strings'. Below I will present a version 
suitable for the purposes of the present work.
The basic point of view we shall adopt here is that the encounter of a 
closed string state with the world-sheet defect (monopole or vortex) 
produces recoil, described by a pair of logarithmic 
deformations~\cite{gurarie} 
corresponding to the collective coordinate 
and velocity of the recoiling D-particle~\cite{kmw,lizzi}, which 
the world-sheet defect is mapped to.

These two operators are slightly relevant~\cite{kmw}, in a 
world-sheet renormalization group sense, with anomalous dimension 
$\Delta = -\frac{\epsilon ^2}{2}$.  
Thus, the recoiling D-particle ceases to be described by a 
conformal theory on the world sheet, 
despite the fact that before the recoil 
the theory was conformal invariant. 
To restore 
conformal invariance one has to invoke 
Liouville dressing~\cite{ddk}, 
thereby increasing the target-space time dimensionality
to $D + 1$.
The Liouville field in this approach 
plays the r\^ole of time, due to the supercriticality~\cite{aben} of the 
central charge ${\cal D}$ of the 
stringy $\sigma$-model, which was critical (conformal) before recoil.
Such a procedure leads to an effective curved space-time $F$-manifold
in $D+1$ dimensions, with signature at least 
as~\footnote{In the 
context of a $\sigma$-model path integral it is convenient to work 
with Euclideanized `times' $X^0$ in the $D$-dimensional space time.
The time $X^0$ is distinct from the Minkowskian signature Liouville time $t$.
In the case of gauge theories, the Euclidean $X^0$
may be thought of as temperature.}
$F=(1,D)$, 
which according to the analysis 
of ~\cite{kanti} is described by a metric of the form: 
\begin{equation}
G_{00}=-1 \,,\, G_{ij}=\delta_{ij} \,,\,
G_{0i}=G_{i0}=f_i(y_i,t)=\epsilon (\epsilon y_i + u_i t)\, ,\,\,i,j=1,...,D
\label{yiotametric}
\end{equation}
where $\epsilon \rightarrow 0^+$, is a regulating parameter, 
related~\cite{kmw} to the world-sheet 
size $L$ via 
\be
\epsilon ^{-2} \sim \eta {\rm ln}(L/a)^2,
\label{epsilon}
\ee
where $\eta =-1$ 
for a Minkowski signature Liouville mode, $t$,  
and $a$ a world-sheet short-distance cut-off. The quantities 
$y_i$ and $u_i$ represent the collective coordinates and velocity of a
$D$-dimensional D(irichlet)-particle. 
In the $\sigma$-model approach of \cite{lizzi,dbrecoil} 
$y_i,u_i$ are viewed as {\it exactly marginal} $\sigma$-model couplings.
Notice that,in view of the identification (\ref{epsilon}), for Minkowskian 
signature Liouville mode $t$, $\epsilon ^2 < 0$. This will be important in
yielding negative curvature anti-de-Sitter space times.  

We recall from the analysis of \cite{kanti} that the components of the 
Ricci tensor for the above (D+1)-dimensional $F$-manifold are: 
\begin{eqnarray}
R_{00}&=& -\frac{1}{(1+\sum_{i=1}^{D} f_i^2)^2}\,
\left(\sum_{i=1}^{D} f_i \frac{\partial f_i}{\partial t}
\right)\,\left[\sum_{j=1}^{D} \frac{\partial f_j}
{\partial y_j} \, \left(1+\sum_{k=1, k\neq j}^{D} f_k^2
\right)\right] \\[3mm]
&+& \frac{1}{(1+\sum_{i=1}^{D} f_i^2)}\,\left[\sum_{i=1}^{D}
\frac{\partial^2 f_i}{\partial y_i \partial t}
\left(1+\sum_{j=1, j\neq i}^{D} f_j^2\right)
\right]\\[5mm]
R_{ii}&=&\frac{1}{(1+\sum_{k=1}^{D} f_k^2)^2}\,
\left\{\,\frac{\partial f_i}{\partial y_i}\,
\left(\sum_{j=1}^{D} f_j\,\frac{\partial f_j}
{\partial t}\right)-(1+\sum_{k=1}^{D} f_k^2)\,
\frac{\partial^2 f_i}{\partial y_i \partial t}\right.\nonumber \\[3mm]
&+& \left. \frac{\partial f_i}{\partial y_i} \left[
\sum_{j=1, j\neq i}^{D}\, \frac{\partial f_j}
{\partial y_j}\,(1+\sum_{k=1, k\neq j}^{D} f_k^2)
\right]\right\} \\[5mm]
R_{0i}&=& \frac{f_i}{(1+\sum_{k=1}^{D} f_k^2)^2}\,
\left\{\frac{\partial f_i}{\partial y_i}\,
\left(\sum_{j=1}^{D} f_j \,\frac{\partial f_j}
{\partial t} \right)-\left(1+\sum_{k=1}^{D} f_k^2\right)\,
\frac{\partial^2 f_i}{\partial y_i \partial t}\right\}\\[5mm]
R_{ij}&=& \frac{1}{(1+\sum_{k=1}^{D} f_k^2)^2}\,
f_i \,f_j\,\frac{\partial f_i}{\partial y_i}\,
\frac{\partial f_j}{\partial y_j} 
\label{Ricci}
\end{eqnarray}

In \cite{kanti} it was argued that due to the abrupt change in the 
surrounding space time at the moment of impact of the closed string state 
$t=0$ with the defect, decoherence will be induced 
for a spectator particle. 
Below we shall consider the {\it asymptotic} case where $t >>0$. 

We shall also restrict ourselves to the case  
where the recoil velocity $u_i \rightarrow 0$.
This case is encountered if the D-particle  
is very heavy, with mass $M \propto 1/g_s$, where 
$g_s \rightarrow 0$ is the (dual) string 
coupling. 
In such a case the closed string state splits into two open ones
`trapped' on the D-particle defect. 
The collective coordinates of the latter, though, are still allowed 
to quantum fluctuate with fluctuations of order~\cite{dbrecoil} 
$\Delta y_i \sim |\epsilon ^2 y_i|$.
From the analysis of \cite{kanti}, then, 
we can infer that, in the $u_i \rightarrow 0$  case, 
the particle creation, which was found to be proportional to $u_i^2$, 
vanishes, and thus 
the asymptotic limit of $t >> 0$ may be treated as a coherent 
quantum string theory, i.e. as a {\it closed} system. 
From the world-sheet point of view~\cite{emnmonop}, 
the very heavy D-particle case 
corresponds 
to a strongly coupled defect. Indeed, the coupling $e$ of the world-sheet defect is related to the (dual) string coupling $g_s$
via 
\be
     e\sqrt{\pi/3} \propto \frac{1}{\sqrt{g_s}} 
\label{duality}
\ee
which implies a world-sheet/target-space strong/weak coupling duality. 
In the case where one views the open world-sheet as 
the area enclosed by a Wilson loop of the gauge theory in target space, then,  
the world-sheet 
defect charge $e$ becomes proportional 
to the target-space gauge theory coupling,
$g_{YM}$. Thus, the above approach allows one to study a {\it strongly coupled
gauge theory} by using a  {\it weakly coupled} (dual) D-string $g_s$. 

We now observe from (\ref{Ricci}) that, in the case of $u_i \rightarrow 0$,  
the only non-vanishing components of the Ricci tensor are: 
\be
R_{ii} \simeq \frac{\partial f_i}{\partial y_i} \left(\sum_{j=1; j\ne i}^{D}
\frac{\partial f_j}{\partial y_j} [1 + {\cal O}(\epsilon ^4)]\right)
\frac{1}{(1 + \sum_{k=1}^{D}f_k^2)^2} \simeq 
\frac{-(D-1)/|\epsilon|^4}{(\frac{1}{|\epsilon|^4} - \sum_{k=1}^{D}|y_i|^2)^2} 
+{\cal O}(\epsilon ^{8}) 
\label{limRicci}
\ee
where we have taken into account (\ref{epsilon}) 
for Minkowskian signature of the Liouville 
mode $t$. Thus, in this limiting case, 
and for large $t >>0$, the Liouville mode decouples, and one is effectively left with a $D$-dimensional manifold. 
We may now write (\ref{limRicci}) as
\be 
   R_{ij}={\cal G}_{ij}R 
\label{newricci}
\ee
with ${\cal G}_{ij}$ a diagonal (dimensionless) 
metric, corresponding to the line element:
\be
      ds^2=\frac{|\epsilon|^{-4}\sum_{i=1}^{D} dy_i^2}{(\frac{1}{|\epsilon|^4} - \sum_{i=1}^{D}|y_i|^2)^2} 
\label{ball}
\ee
which is the metric describing the {\it interior} 
of a $D$-dimensional ball, 
depicted in figure 1. This is the Euclideanized version of an 
anti-de-Sitter space time~\cite{witten}. In its Minkowski version,
one can easily check that the curvature corresponding to (\ref{ball}) 
is {\it constant} and {\it negative}, $R = -4D(D-1)|\epsilon| ^4$.

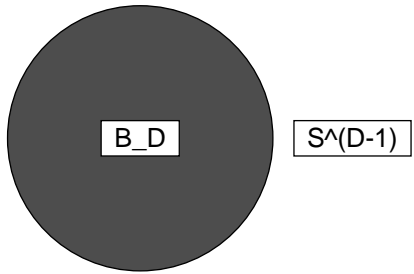
\begin{figure}
\begin{center} 
\begin{picture}(120,120)(70,0)
\SetPFont{Helvetica}{10}
\SetScale{1}
\GCirc(100,100){50}{0.3}
\GText(100,100){1}{B_D}
\GText(180,100){1}{S^(D-1)}
\end{picture}
\end{center}
\caption{Schematic representation of the 
$D$-dimensional (Euclideanized) anti de Sitter space time, arising in our
Liouville approach to a recoil D-particle, as the interior 
of a $D$-dimensional ball $B_D$; the boundary of the anti-de-Sitter
is a $D-1$ dimensional conformal sphere $S^{D-1}$ 
(compactified Minkowski space time).
Any field theory defined on $S^{D-1}$ may be viewed as corresponding uniquely 
to a D-brane supergravity theory in the bulk of the $AdS_D$.} 
\label{fig1}
\end{figure}

The interesting property of the space time (\ref{ball}) lies on the fact that 
there exist a coordinate singularity in the metric (\ref{ball}) 
at $\sum _{i=1}^D|y_i|^2=|\epsilon|^{-4}$, which prevents a naive extension 
of the open ball $B_D$ to the {\it closed} ball 
${\overline B}_D$, with boundary the sphere $S^{D-1}$. The metric that extends 
to ${\overline B}_D$ is provided by a {\it conformal} 
transformation of (\ref{ball})~\cite{witten}: 
\be
    d {\tilde s}^2=f^2 ds^2 
\label{conformal}
\ee
Choosing, say, 
$f=\frac{1}{|\epsilon |^{-4} - \sum_{i=1}^D|y_i|^2}$, 
results in $d{\tilde s}^2$ being associated with 
the metric on a sphere $S^{D-1}$ of radius 
$\frac{1}{|\epsilon|^2}$.  In general $f$ may be changed by any conformal transformation, thereby leading to the appearance of a conformal invariant 
(Euclidean) $S^{D-1}$ space as the {\it boundary} of an 
anti-De-Sitter space time, whose metric is invariant under the Lorentz group 
SO(1,D). 

This approach to AdS space time and their connection to conformal 
invariant 
Minkowskian space times is due to Witten's elaboration~\cite{witten} on the 
conjecture by Maldacena~\cite{malda} that certain large N-limit conformal field theories in $d$ dimensions can be described in terms of supergravity 
(or string theory) on the product of a $(d+1)$ dimensional AdS space with a compact manifold.  In our approach above we have shown how AdS string backgrounds can arise naturally in the Liouville $\sigma$-model 
approach to D-brane 
recoil advocated in \cite{dbrecoil}. 
The discussion then of refs. \cite{witten},\cite{malda} 
applies to the resulting anti de Sitter space times~\footnote{It should be 
stressed, however, that our approach, as discussed in \cite{kanti}, 
is much more general. The AdS space time was obtained in the limit of zero recoil velocity $u_i$, whilst the inclusion of such velocity corrections 
complicates the asymptotic definition of a local quantum field theory
on the boundary of our space time, due to induced decoherence as a result of quantum recoil degrees of freedom.}.   
The important property 
of Anti-de-Sitter (AdS) space times is the
existence of powerful theorems which imply 
that if a classical 
field theory is specified in the boundary of the space, then there 
{\it are} unique extensions to the bulk~\cite{witten,lee}. 
This is the main point of the 
{\it holographic} nature of field/string theories in anti-de-Sitter
space times,  
in the sense that all the information about the bulk theory of an anti-de-Sitter space is `stored' in its 
boundary. By the Gauge Theory/Conformal-field-theory 
correspondence of \cite{malda,witten}, then, 
information 
about quantum aspects of gauge theories on the boundary of $AdS$, are encoded in classical geometry in the bulk of the $AdS$ space.

We may now interpret the Ricci tensor (\ref{limRicci}) as corresponding 
to the 
low-energy- ${\cal O}(\alpha ')$, $\alpha ' << 1$- 
$\beta$ function of a one-string-loop corrected $\sigma$-model, propagating in this background. Notice that the one-string-loop correction is necessary so as to yield the constant-curvature nature of the background in a way compatible 
with the conformal invariance~\cite{fischler}. Indeed, in a perturbative expansion in terms of $\alpha '$, the ${\cal O}(\alpha ')$ $\sigma$-model $\beta$-function is proportional to the Ricci tensor $R_{ij}$ alone~\cite{fradkin}.
The graviton background conformal invariance conditions, imply the relation 
\be
    \beta ^G_{ij} = R_{ij}=\nabla _i \partial _j \Phi 
\label{betafunction}
\ee
with $\Phi $ a dilaton field. 
At first sight, it seems  
that the Ricci tensor (\ref{newricci}), describing  a constant curvature 
space time, cannot be a consistent string background,
compatible with conformal invariance (\ref{betafunction}),  to order 
$\alpha '$. However, as shown in \cite{fischler}, this conclusion 
is false. Indeed, if 
one includes higher-string loop corrections, then 
one obtains  an induced (target-space) 
cosmological constant, corresponding to a dilaton 
tadpole, which renders the constant-curvature backgrounds consistent 
with the conformal-invariance conditions,
corrected by  
string loops.    
The mechanism of \cite{fischler} concerns one loop corrections, which in our case
are sufficient, in view of the smallness of the 
(dual) 
string coupling $g_s << 1$. 
According to \cite{fischler}, 
there are {\it zero momentum} target space dilatons, non-propagating delocalized modes, which fail to decouple from the one loop partition function in pinched world-sheet loops (see figure 2).
Such modes yield non-zero constant {\it dilaton} tadpoles, $J$.

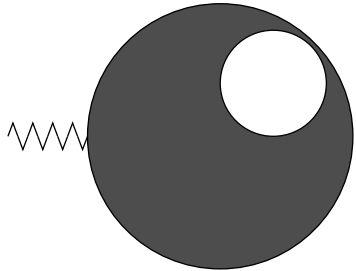
\begin{figure}
\begin{center} 
\begin{picture}(120,120)(70,0)
\SetPFont{Helvetica}{10}
\SetScale{1}
\GCirc(100,100){50}{0.3}
\GCirc(120,120){20}{1}
\ZigZag(20,100)(50,100){5}{4}
\end{picture}
\end{center}
\caption{Schematic Representation of 
dilaton (zigzag lines) tadpoles on pinched world-sheet surfaces,
which are essential for compatibility of anti-de-Sitter target-space 
backgrounds with 
$\sigma$-model 
conformal invariance conditions.} 
\label{fig2}
\end{figure}

Such pinched loops lead to extra world-sheet divergences, 
whose regularization   
in lower genus (tree level) world-sheet surfaces induces 
corrections to the graviton $\beta$ functions 
of the form 
\be
   \beta_{ij}^{G,1-string-loop}=\beta _{ij}^{G,tree-level} + 
g_sG_{ij}J, \qquad J={\rm zero~momentum~dilaton~tadpole}
\label{tadpoledil}
\ee
thereby leading to a constant curvature space time, with cosmological 
constant 
$g_sJ$,  
as a consistent solution 
of the string-loop corrected conformal invariance conditions. 
In our case above, $g_sJ \propto -|\epsilon|^4$. 

Notice that the one-string-loop correction is consistent with the fact that 
in the context of our original $\sigma$-model, the recoil operators
are themselves viewed as arising from one string loop corrections 
in the propagating string matter states (see fig. 3a), 
due to the cancellation of extra world-sheet divergences arising from pinched 
world-sheet loops (see figure 3b), 
in the presence of Dirichlet 
boundary conditions~\cite{periwal,dbrecoil}. 
Also notice that the 
the fact that the solution is 
of anti-De-Sitter type, allows extension of these ideas to  
target-space supergravity theories, given that 
that local supersymmetry
in a space of constant curvature for $D > 2$  {\it requires} 
that the space time is of Anti-de-Sitter type~\cite{salam}. 

\begin{centering} 
\begin{figure}[htb]
\epsfxsize=3in
\centerline{\epsffile{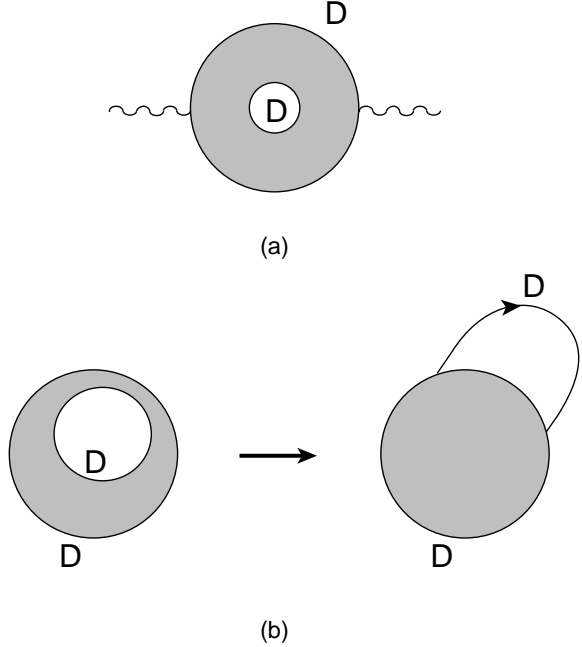}}
\vspace{1cm}
\caption[] { (a) : World-sheet
annulus diagram 
for the leading quantum correction to the 
propagation of a string state in a $D$-brane background, and 
(b) the pinched annulus configuration which is the dominant divergent 
contribution to the quantum recoil.}
\label{fig3}
\end{figure} 
\end{centering}

The above discussion has indicated the appearance of an 
anti-de-Sitter (target) space time, 
due to the recoil of a very heave D-particle.
As well understood by now~\cite{dbranes}, a
single D-particle corresponds 
to a (compact) $U(1)$ gauge theory~\cite{dbranes}, whilst 
$m$ D-particles, or $m$ parallel non-interacting branes,  
correspond to a $U(m)$ gauge symmetry~\cite{dbranes,lms}.
In the latter case, 
the formal extension of the above approach is 
computationally  
non trivial. 
The formalism is complicated due 
due to technicalities involved in 
the path ordering of the non-Abelian Wilson loops~\cite{lms}.
However, there appear to be no conceptual problems with such an extension. 
What we conjecture as happening, which we shall justify in  
this article, is that, in the {\it non Abelian} case, the topology of the 
anti-de-Sitter space time may be more involved. In particular, 
in the five-dimensional case, of relevance to QCD,
the above-described Liouville approach 
to $D$-particle recoil
may lead, in certain cases, 
to AdS space-times involving  
Black Holes~\cite{page}, 
thereby generalizing non-trivially 
the $B_{d+1}$ (\ref{ball}), 
encountered in the 
single-D-particle case above.

\subsection{Liouville-String Approach to Four-Dimensional Gauge Theories}

Let us discuss now 
the connection of D-brane recoil 
(viewed as quantum fluctuations of a heavy D-particle in 
a (dual) space time) 
to a Gauge Theory 
at {\it finite temperatures}, 
formulated in {\it real time} formalism. 
In that case, the gauge theory lives in a five-dimensional space time
manifold, four space-time Minkowskian, with a time $t$, and a compactified 
time direction, playing the r\^ole of temperature.

Consider first the Matsubara
formalism, where the finite-temperature field theory 
is obtained by compactifying the time direction to a circle $S^1$ 
of period the inverse temperature $\beta$. 
We consider the case where the space is flat.
In this case, 
the string theory, 
living on the two-dimensional open string, whose boundary 
is the Wilson loop of the Gauge theory, may be taken to be a 
non-critical Liouville string theory with $\sigma$-model action: 
\be
{\cal S}_\sigma = \int_{\Sigma (C)} \{ 
(\partial \phi )^2 + Q \phi R^{(2)} + \Phi (X) R^{(2)} 
+ G_{ij} (X) \partial X^i \partial X^j 
+ V_v(\phi ) + V_m (\phi) + \dots \}
\label{ncsg}
\ee
where $X^i, i=1, \dots 4$, span a manifold of topology $R^3 \otimes S^1$,  
$\phi $ denotes the Liouville mode, and the dots 
denote appropriate supersymmetrizations or other deformations,
relevant in the description of the Wilson loop, such as 
antisymmetric tensors etc~\cite{polyakov}.
$Q$ is related to the 
matter central charge deficit in a standard way~\cite{aben,ddk}. In 
our case, we initially take 
the matter central charge to be $c=4$, 
coinciding with the dimensionality of the space time 
manifold $X$. 
There are world-sheet vortices 
and monopoles in such a picture, since the Liouville theory is in the 
dangerous region, $1 < c < 9$, 
so the deformations $V_v,V_m$ should be thought of 
as the analogue of the Liouville `cosmological constant' operators 
in the $c<1$ case. These deformations are given~\cite{ovrut} by 
(\ref{susyvortex}),(\ref{susymonopole}), respectively. The 
Liouville mode $\phi $ 
is `space-like' in this case. Moreover, if one views the Liouville field $\phi$
as a local scale on the world-sheet~\cite{emn}, then the space-time metric
should be renormalized by $\phi$, which leads to curved
five-dimensional manifolds,  
$G_{AB}(X,\phi)$, $A,B=1,\dots 5$. 
Thus, a four-dimensional (three spatial dimensions and a 
temperature) string is thereby  
dressed with an extra 
space-like target dimension, 
in order to become critical and thus quantized
self-consistently.

We next come to examine the r\^ole of a Minkowskian time in such a formalism,
which turns out to be a second (time-like) Liouville field, 
that determines 
the consistent string ($\sigma$-model) 
backgrounds for the five-dimensional metric $G_{AB}$.  
Including the time $t$ as a dynamical field in the $\sigma$-model 
(\ref{ncsg}), corresponds, 
from the point of view of the original gauge theory, 
to a `real time' approach to the finite temperature field theory. 
Let us see now how we can incorporate the field $t$ consistently. 

To this end, we first recall that 
a Wilson loop may be thought of as describing 
a heavy test particle propagating along the loop. 
At the beginning of time, $t=0$, the particle starts moving, which
causes sufficient distortion of the surrounding space time,
described in the dual string picture by the recoiling $D$-particle.
In terms of the `Wilson-loop 
$\leftarrow\rightarrow$ string correspondence', discussed above,
the situation 
may be thought of as corresponding to placing 
Dirichlet boundary 
conditions on the boundaries of the world-sheet disc describing the two-dimensional extension of the Loop curve $C$ (Stokes' theorem)~\footnote{Such boundary 
conditions
may arise by viewing $T$ duality as a {\it canonical} 
transformation of the string world sheet~\cite{dorn,dbrecoil,am}.}. 
Consider the case where the $D$-particles are 
quite heavy, so any velocities $u_i \rightarrow 0$.
Then, in a weakly dual string $\sigma$-model picture
the deformation is 
equivalent to the 
position quantum 
fluctuations of 
the D-particles, 
appearing at time $t=0$, and it is given by  
the recoil operator~\cite{kmw}   
\be
      y_i \int _{\partial \Sigma = C} \Theta _\epsilon (t) \partial _n X^i 
= y_i \int _{\Sigma (C)} \partial _\alpha (\Theta _\epsilon (t) \partial ^\alpha X^i) 
\label{recpos}
\ee
where we have used Stokes theorem to write the boundary recoil operator as a bulk operator~\cite{dbrecoil}. Notice that, 
the size $(L/a)^2$ (where 
$a$ a world-sheet cut-off), of the world-sheet disc is to be identified 
with the size of the 
quark-anti quark loop, so the $\epsilon ^2 \sim 1/{\rm ln}(L/a )^2
\rightarrow 0$ is consistent with considering large loops, 
relevant for the confining aspects of the Gauge theory (!).
The collective coordinates for the position of the D-particles $y_i$
describe the space-time coordinates of the gauge theory, or equivalently
the target space 
of the non-critical string theory living on the world-surface 
surrounding the Wilson loop.

Since the recoil operator (\ref{recpos}) 
has small but negative anomalous dimension~\cite{kmw} $-\epsilon ^2/2$,
one may dress this theory with a time-like Liouville field, which
introduces an extra time-like dimension ${\varphi }$,
since, as mentioned previously,  
the five-dimensional $\sigma$-model is 
assumed critical, before the recoil, 
as a result of the (space-like) Liouville dressing.
By identifying $\varphi $ with $t$, as in \cite{dbrecoil}, 
one, then, obtains 
a $\sigma$-model in {\it six} dimensions, with signature (1,5). 
This six-target-dimensional 
Liouville dressed theory is described by a target-space metric around 
the D-particle of the form
(\ref{yiotametric}), leading to a target-space time Ricci curvature
of the form  (\ref{Ricci}). 
In such a case, 
the $t$-Liouville-dressed couplings $y_{i,L}$ are given to lowest order by: 
\be
    y_{i,L} =\epsilon y_i e^{\epsilon \varphi } \qquad ; 
\qquad \varphi \equiv t 
\label{yr}
\ee
and the approach of \cite{dbrecoil} applies 
straightforwardly to this case, 
since the only source of non-criticality is the operator (\ref{recpos}).

To find the consistent geometry of the five-dimensional Euclidean 
manifold,
one should then follow 
the steps outlined in the previous section, leading to 
(\ref{yiotametric}), which is achieved 
straightforwardly  by partial 
differentiating the bulk vertex operators 
(\ref{recpos}), as in \cite{dbrecoil}. 
Then, in 
the limit of a heavy D-particle (weakly coupled string $g_s << 1$) 
the time (extra time - like Liouville) coordinate $t$ 
decouples, and the distortion of space time may  
be described by anti-de-Sitter geometries 
for the Euclidean $5$-dimensional space. 

Recalling  that 
the latter theory has a 
`matter central charge' $4$, and hence a
space-like 
Liouville mode, then, 
according to our discussion in the previous section, 
we observe that 
the $\sigma$-model associated with this AdS background 
has a similar world-sheet phase diagram 
for the vortices and monopoles with the one 
described in (\ref{regions2}),(\ref{regions3}).
The various phases are described by non-critical strings
with varying matter central charges, ${\cal D}$, 
related to the temperature
via (\ref{susybeta}).  
In such a non-critical string picture, 
the deficit of the target-space dimensionality (D=5) is compensated 
by condensates of the background dilaton (and possibly other) 
fields. The r\^ole of the dilaton condensate  may
be associated with condensates of glueball fields, related to 
the scale anomaly 
in effective Lagrangian descriptions of ordinary $QCD$~\cite{ellis}. 
We hope to come back to a detailed discussion of these issues in 
a forthcoming publication~\cite{em}.

As we shall discuss below,
there is an elegant interpretation of this non-critical string 
approach  
to the real-time formalism of finite-temperature 
gauge theories, in terms of thermodynamical 
properties of a gas of (classical) AdS 
black holes~\cite{page}, as a result 
of the AdS/CFT correspondence~\cite{malda,witten}.
The above-described world-sheet defect description, allows 
an identification  
of the pertinent `Liouville temperature'
(\ref{susybeta})
with the temperature of the (macroscopic) $AdS_5$ 
~\cite{page}, whose boundary is the conformal 
space time, on which  
the (conformal) gauge theory in question lives. In this way  
one may obtain 
non-perturbative information on 
the phases of the gauge theory, which can be identified with the 
phases (\ref{regions2}) of the corresponding 
Liouville theory. However, in the present work, 
we shall also  point out 
an alternative 
possibility, according to which 
the pertinent $AdS_5$ Black Hole structures, are  
not associated with {\it classical macroscopic geometries} 
as in \cite{witten}, but with 
{\it microscopic fluctuations}, whose size is set by the 
UV cutoff of the low-energy theory. The latter induce an area law
for large Wilson loops, but with string tension considerably weaker 
than the previous macroscopic $AdS_5$ geometries. This latter case 
is probably 
relevant for the appearance of the fundamental string tension
in grand-unified models of strings.

\section{Coupling of Scale Invariant (Super)String theories to
Gauge Theories}

As a prelude to our results, we consider it as instructive to review first 
briefly 
the work 
of \cite{awada}, according to 
which certain flat-target-space scale invariant 
Green-Schwarz  
superstring theories 
can be shown to be 
equivalent to 
Abelian gauge theories.
This equivalence should be understood 
in the sense of relations of the form:
\be 
<W(C)>~\sim~e^{iS_\sigma }
\label{awadaabel}
\ee
with $S_\sigma $ a world-sheet 
action, expressing area of the Wilson loop in some sense, 
and $W(C)$ some combination of  
observables, expressed in terms of chiral currents
$e^{\int J.A}$, which 
goes beyond the standard Wilson loops. The action $S$ becomes a standard 
world-sheet action   
upon appropriate dynamical generation of a string scale
due to some sort of 
condensation. In \cite{awada},  
Supersymmetry was argued to be {\it essential}  
for such an equivalence~\cite{awada}.
It is the point of our work  to argue that if 
monopoles or vortices 
are present on the world sheet, then  the supersymmetry becomes 
unnecessary, and in fact one may get a natural 
mechanism for exhibiting the confining aspects of the finite-temperature 
gauge theory,
even in non-supersymmetric (but still conformal) cases.

The analysis of ref. \cite{awada} was performed 
for four-dimensional target spaces, but it can be 
straightforwardly generalized to higher dimensionalities as well. 
For reasons of concreteness
and calculational simplicity,
we shall review below the analysis in the 
four dimensional case, where the gauge field theory coupling is 
{\it dimensionless}. 
We shall first review the (supersymmetric) case where world-sheet
defects have been ignored~\cite{awada}. We shall comment on the non-trivial 
r\^ole of our defects in the next section. 

Let us consider 
an 
Abelian Supersymmetric Gauge Theory, described by a standard Maxwell 
Lagrangian in a superfield form. 
The connection with string theory follows from standard arguments, 
by considering the Stokes theorem 
on a two-dimensional surface, $\Sigma$, 
whose boundary is the loop $C$ 
in question. If one parametrizes the curve $C$ by $\lambda$, then he    
may write the exponent of the Wilson loop as 
\be
     S_{int} =ie \int _{C} d\tau A( X(\tau)) \frac{\partial }{\partial \tau } X(\tau) 
\label{wilsonexp}
\ee
and by applying  Stokes' theorem 
\be
   S_{int} =\frac{ie}{2}\int _{\Sigma (C)} d^2\sigma \epsilon^{ab}
F_{\alpha\beta}, 
\qquad a, b =1,2,
\label{stokes} 
\ee
where $X^M$, $M=1, \dots D$, is a $D$-dimensional 
space-time coordinate for the gauge theory; we denote 
by lower-case Latin indices 
the two-dimensional coordinates of the surface $\Sigma$, 
which plays the r\^ole of the world-sheet of the string, equivalent to the gauge 
theory in question~\cite{polyakov}. The quantity 
$F_{ab} = \partial_a X^M \partial _b X^N F_{NM} =
\partial _a A_b - \partial _b A_a $
is then the pull back of the Maxwell tensor on the `world-sheet' 
$\Sigma$, with $A_a$ the corresponding projection of the gauge field 
on $\Sigma$ (c.f. (\ref{pullback2}): 
\be
A_a =v_a^M  A_M \qquad ; \qquad v_\alpha ^M \equiv \partial _a X^M , 
a=1,2~; M+1, \dots D(=4).
\label{pullback3}
\ee
From a two-dimensional 
view-point, 
the term looks like a Chern-Simons term for a two-dimensional 
gauge theory on $\Sigma$, bounded by the loop $C$. 

The novel observation of \cite{awada}
is the possibility, 
in a supersymmetric gauge theory,  
of constructing, 
a {\it second} superstring-like 
observable, in addition to the Wilson loop, which again is defined on the 
two surface $\Sigma$, and is consistent with all the symmetries of the theory.
The second observable is easily understood in the two-dimensional 
superfield formalism 
\be
Z^{{\cal A}} 
\equiv (X^M, \theta ^m, \theta ^{{\dot m}})
\label{sf}
\ee
The pull-back basis $v_a ^M$ in (\ref{pullback3}) is now extended
to $v_a^{{\cal A}} = E^{{\cal A}}_{{\cal B}} \partial _a z^{{\cal B}}$,
with 
the following components~\cite{awada}:  
\bea
&~&v_a^{\alpha{\dot \alpha}} =\partial _z x^{\alpha{\dot \alpha}}
-\frac{i}{2}\left(\theta^\alpha (\sigma) \partial _a \theta ^{{\dot \alpha}} (\sigma)
+ \theta^{{\dot \alpha}} (\sigma) \partial _a \theta ^{\alpha} (\sigma)\right)   
\nn \\
&~& v_a^\alpha = \partial _a \theta ^\alpha (\sigma) \nn \\
&~& v_a^{{\dot \alpha}} = \partial _a \theta ^{{\dot \alpha }} (\sigma)   
\label{superspace}
\eea
in the standard notation~\cite{superspace}, where Greek dotted and undotted 
indices denote superspace components, with   $x^{\alpha{\dot \alpha}}
\equiv X^M$, etc. 

Following \cite{awada} we now define: 
\bea 
&~& C_{ab}^{\alpha\beta} \equiv \frac{i}{2} 
v_{a{\dot \beta}}^{(\alpha}
v_{b}^{\beta){\dot \beta}}
\nn \\
&~&       C_{ab}^\alpha \equiv v_a^{\alpha{\dot \alpha}}v_{b{\dot \alpha}}
\nn \\
&~& {\rm with~the~properties} \qquad 
C^\alpha =\epsilon^{ab}C_{ab}^\alpha , \qquad C^{\alpha\beta}=\epsilon^{ab}C_{ab}^{\alpha\beta}
\label{Cs}
\eea
with similar relations holding for appropriately  defined 
dotted components of $C$, which can be found in \cite{awada}. 
Notice that $C^{\alpha}$ {\it vanishes in the absence
of supersymmetry}, whilst $C^{\alpha\beta}$ exists even 
in non supersymmetric gauge theories as well.  
In the presence of defects, as we shall later, this is 
no longer the case of course, and $C^\alpha$ can have non-supersymmetric 
remnants. 

The supersymmetric Wilson loop, expressing the interaction between the 
superparticle and a supersymmetric gauge theory in the approach of \cite{awada},  
is now expressed in terms of $C's$ as: 
\bea
 &~& W(C)=e^{S_{int}^{(1)}} \qquad 
S_{int}^{(1)} \equiv \frac{i}{2} \int _{\Sigma (C)} d^2\sigma \epsilon^{ab}
{\cal F}_{ab} \nn \\
&~&{\cal F}_{ab} \equiv \epsilon_{ab}\{\frac{1}{2}C^{\alpha\beta}(\sigma)
D_\alpha W_\beta (x(\sigma),\theta (\sigma)) + \nn \\
&~& C^\alpha (\sigma) W_\alpha
(x(\sigma),\theta (\sigma)) + h.c. \} 
\label{susywilson}
\eea
where $W_\alpha (x(\sigma),\theta (\sigma))$ is the chiral superfield 
of the  supersymmetric Abelian gauge theory. 
The exponent $S_{int}^{(1)}$ 
 clearly reduces to the standard expression (\ref{wilsonexp}) in non-supersymmetric cases.

According to \cite{awada}, 
there is a second superstring-like observable, as noted in \cite{awada},
defined on the world-sheet surface $\Sigma$,
$\Psi (\Sigma )$, 
which is constructed out of the $C_{ab}^\alpha $ components, and hence -
in the absence of world-sheet defects - 
{\it exists only in supersymmetric gauge theories}, as mentioned previously:
\be
 \Psi (\Sigma ) \equiv e^{iS_{int}^{(2)}}, \qquad 
S_{int}^{(2)} \equiv \kappa \int _{\Sigma (C)} d^2\sigma 
\sqrt{-\gamma}\gamma ^{ab}C_{ab}^\alpha (\sigma) 
W_\alpha (x(\sigma), \theta (\sigma)) + h.c.
\label{second}
\ee
where $\gamma ^{ab}$ is the metric on $\Sigma$. This term,
unlike the standard Wilson loop, 
{\it is not} 
a total world-sheet derivative, and, therefore, lives in the bulk of the 
world-sheet $\Sigma$, and thus depends on the metric $\gamma $. 
The coupling constant $\kappa$ 
is defined as an independent coupling classically. However, one expects 
that 
quantum effects will relate it to the gauge coupling 
constant $e$. We shall come to this point later. 
In \cite{awada} this second observable has been expressed in terms of `chiral' currents on $\Sigma$, located at the string source:
\bea
&~&   S_{int}^{(2)} =\int d^6Z \left({\cal J}^\alpha 
W_\alpha + h.c. \right), \nn \\
&~& {\cal J}^\alpha \equiv \kappa \int _{\Sigma (C)} d^2\sigma \sqrt{-\gamma}\gamma ^{ab}C_{ab}^\alpha (\sigma) \delta ^{(6)} (Z-Z(\sigma)),  
\nn \\
&~& \delta ^{(6)} (Z-Z(\sigma)) =\delta ^{(4)} (Z-Z(\sigma))\left(\theta - \theta (\sigma)\right)^2 
\label{chiral}
\eea
Upon integrating out the gauge field components in (\ref{second}), 
i.e. considering the vacuum expectation value (vev): 
$<\Psi (\Sigma )>$, with $< \dots >$ denoting 
averaging with respect to the Maxwell action for the gauge field, 
the authors of \cite{awada} 
have obtained a superstring-like action, which is 
{\it scale invariant} in target space, as well as on the world-sheet:
\bea
 &~&   <\Psi (\sigma )>_{Maxwell} =e^{S_0 + S_1}  \nn \\
&~& S_0 \equiv \frac{\kappa _0^2}{16\pi} \int _{\Sigma } d^2 \sigma 
\sqrt{-\gamma}\gamma ^{ab}v_a^Mv_b^N\sigma _M\sigma _N, \nn \\ 
&~& S_1 \equiv \frac{\kappa _1^2}{4\pi} \int _{\Sigma (C)} 
\sqrt{-\gamma} \gamma ^{ab}v_a^M v_b^N \eta_{MN} \sigma^K\sigma_K
\label{gaugefield}
\eea
where upper case latin indices denote target space indices ($M,N,K=1, \dots D(=4$)), 
and 
\bea
&~& v_a^M \equiv \partial_a X^M(\sigma) - 
i{\overline \theta}^m(\sigma)\Gamma ^M \partial _a \theta _m  (\sigma) 
\nn \\
&~&    \sigma^M=\frac{\sqrt{-\gamma}\epsilon^{ab}}{\sqrt{{\rm det}G}}\partial_a v_b^M, \quad G_{ab} \equiv v_a^Mv_b^N\eta_{MN} 
\label{fourcomp}
\eea
in standard four-component notation in target space ($m$ are spinor indices, 
and $\Gamma ^M$ are Dirac four-dimensional matrices). 
The (dimensionless) coupling constants $\kappa _{0,1}$ appear arbitrary 
at a classical level, but one expects them to be related
(proportional) to the (dimensionless) 
gauge coupling $e$, in the quantum theory.

The important observation in \cite{awada} was the fact that the 
world-sheet action $S_2$ (\ref{second}) resembles the
classical Green-Schwarz superstring action in flat four-dimensional 
target space, provided that condensation occurs for the `composite field' 
\be
\Phi \equiv \sigma^M \sigma_M,\qquad M=1, \dots D(=4), 
\label{composite}
\ee
in such a way that 
\be
\frac{\kappa _1^2}{4\pi}<\Phi>=\mu_{string~tension}
\label{tension}
\ee
In this way a dimensionful scale (string tension) could be obtained from a 
gauge theory without dimensionful parameters.
The physically interesting question 
is what causes this condensation, which presumably 
is responsible for 
a `spontaneous breaking' of 
the scale invariance of four dimensional string theory. 

In the next section we shall present a scenario on the formation of 
such a condensate, which is an extension of the ideas of \cite{witten}. 
We shall 
associate the second observable (\ref{second}) 
of \cite{awada} to the condensation 
of non-trivial world-sheet defects. 
The latter, in turn, will  
be associated with 
topology changes in the five-dimensional Liouville $AdS$ space time,
argued previously to 
arise in the non-critical string approach to the four-dimensional 
gauge theory, described in section 3.  
In such a picture, 
the standard Wilson loop $W(C)$ will exhibit the area law
in the confining (low-temperature) phase, with string tension 
(\ref{tension}), which can be 
related to the radius of the $AdS_5$
space time. Notice that discussions in the recent literature on 
non-perturbative prescriptions for the computation 
of Wilson loops~\cite{wilsoncomp} apply to our case 
as well, provided one takes proper account of the 
Liouville dynamics, and the non-trivial world-sheet topology 
due to the presence of the defects. We hope to give detailed reports 
on such delicate matters
in the future.  At the moment we note that 
the condensate of $\Phi$ (\ref{composite}) 
may be viewed as being related to a vacuum expectation value 
of a target-space dilaton field, which in this approach may be related
to scalar glueballs, appearing due to scale anomalies in standard 
treatments of effective Lagrangians for $QCD$~\cite{ellis}. 
We plan to discuss these issues elsewhere.

\section{World-Sheet Monopoles and Quark Confinement} 

\subsection{A Small Digression on Anti-De-Sitter Black Holes} 

One possible mechanism for the appearance of the condensate 
(\ref{tension})
would be the one 
associated with 
four-dimensional 
quantum gravity effects at Planckian scales. 
In such a case the resulting string tension would probably describe fundamental strings, responsible for the unification in higher dimensional string 
models~\footnote{Notice that if one 
defines the composite field $\Phi$ as above, 
then one may extend the above analysis to higher dimensional 
cases by simply contracting the integrand of (\ref{second}) 
with appropriate powers of $\Phi $ so as to make the coupling constant 
$\kappa$ {\it dimensionless}. This can be understood as the definition of the 
superstring observable $\Psi $ in higher than four target space dimensions.
It is important 
to note that {\it the dimensionality of the coupling constant} $\kappa$
{\it is the same as that of the gauge theory coupling} $e$ {\it only 
for space-time dimensionality four, six and ten}, where supersymmetric theories exist as well.}. 

A more elegant scenario, which is probably more relevant 
for our 
long-distance confining gauge theory physics, 
may be based on the recent 
interpretation~\cite{witten} of confinement in pure glue 
non-Abelian gauge theories, by means of a 
holographic principle encoding 
confinement quantum physics to classical 
geometries of (uncompactified) five-dimensional anti-De-Sitter space times. 
In such a picture,  
the four-dimensional conformally 
invariant Minkowski space time is 
viewed as 
the boundary of a 
five-dimensional Anti-de-Sitter 
space time, in the spirit of \cite{malda,witten}. 
{\it Macroscopic } Black Holes   
defined in anti-de-Sitter space times 
disappear at temperatures $T_0$, as we go down in temperature 
from a high-temperature phase. 
Notice that Macroscopic Black holes in anti-de-Sitter 
space times have been found to posses 
well-defined thermodynamical properties~\cite{page}. 
Indeed, the analysis of \cite{page} has shown that 
there exists a critical temperature, $T_0$, above which black holes 
can be thermodynamically stable. Below $T_0$ only radiation-dominated
universes exist. 
Witten~\cite{witten}, has associated 
this temperature to a confining-deconfining phase 
transition for quarks, and stressed the fact that 
for spatial Wilson loops the are law and the associated string tension is obtained only for finite temperature field theory; this is due to the 
conformal invariance of the zero-temperature four-dimensional gauge theory, 
whose Renormalization Group $\beta$ function vanishes identically. 

It is instructive for our purposes to review briefly the 
relevant  
properties   
of Anti-De-Sitter space times~\cite{page,witten}. 
For concreteness we shall explicitly describe
the case of \cite{page}, which is $AdS_4$. The generalization 
to $AdS_d$, for general $d$ is straightforward~\cite{witten}. 

The (Minkowskian signature ) AdS 
Schwarzschild Black hole solution of \cite{page} 
corresponds to a line metric element of the form: 
\be
   ds^2 = -V (dt)^2 + V^{-1} (dr)^2 + r^2 d\Omega ^2
 \label{adsbh}
\ee
with $d\Omega ^2$ the line element on a round two-sphere, and 
$r$ the radial coordinate of $AdS$, $t$ is the time coordinate. 

The Anti-de-Sitter Black Hole space time, which is obtained as a 
consistent {\it classical} solution of Einstein's equations, 
in a space with cosmological constant $\Lambda <0$, 
becomes a smooth geometry if one compactifies the time direction, 
to a special period $\beta$, defining a 
temperature for the AdS Black Holes
\be 
       \beta = \frac{4\pi b^2 r_+}{b^2 + 3r_+^2}, 
\qquad b=\sqrt{-\frac{3}{\Lambda}}
\label{hptempr}
\ee
where 
$r_+$ is the maximum of the roots of the equation 
\be
V \equiv  1- \frac{2M}{m_p^2r} + \frac{r^2}{b^2}, 
\label{V}
\ee
where $m_p$ is the $AdS$ space Planck mass.   
The Euclidean version of the space time has the topology of 
\be
     X_2 =B^2 \times S^{n-1}~~~({\rm n=3~for~ref.~\cite{page}})
\label{x2}
\ee

According to the analysis of \cite{page}, 
there are three critical temperatures in the AdS black hole system:
(i) at the first one $T_0$ 
the specific heat of a gas of 
Black Holes changes sign at a critical temperature, $T_0$. 
The critical temperature is the one corresponding to the maximum of $\beta $ 
(\ref{hptempr}) :
\be
      T_0 =(2\pi)^{-1}\sqrt{3}b^{-1} 
\label{crit}
\ee
For $T < T_0$ there is only radiation, and the topology of 
the finite-temperature space time 
is described by:
\be 
   X_1 = B^n \times S^1~~~({\rm n=3~for~ref.~\cite{page}})
\label{x1}
\ee 
The line element for this metric, without Black Holes, 
is given by:
\be
ds^2_{x1}=(1 + \frac{r^2}{b^2})
dt^2 + \frac{1}{1 + \frac{r^2}{b^2}} dr^2 + r^2 d\Omega ^2 
\label{dsx1}
\ee
(ii) For temperatures above $T_0$ 
the topology of the space time 
changes to $(\ref{x2})$ to include black holes, but 
in the region $T_0 < T < T_1$, where: 
\be
T_1 = \frac{1}{\pi}b^{-1} 
\label{t1}
\ee
the free energy of the black hole is positive, so the black hole 
would have the tendency to evaporate. \\
(iii) For $T > T_1$,   
the free energy of the configuration with the black hole 
and thermal radiation is lower than the corresponding 
configuration with just thermal radiation,
so the radiation would just tunnel to black holes,  
whilst \\
(iv) for temperatures greater than a third value $T > T_2$ : 
\be
T_2 = (m_p^2)^{1/4}(3)^{1/4}b^{-1/2}
\label{t2}
\ee
there is no equilibrium
configuration without a black hole. 

The {\it topology change} in the five-dimensional AdS space time 
from $X_1$ (\ref{x1}) to $X_2$ (\ref{x2}), at $T=T_0$, prompted 
Witten~\cite{witten} 
to conjecture, based on the CFT/AdS correspondence~\cite{malda,witten},
that there exists a connection 
of this phase transition, which in the case of a gas of 
AdS Black Holes is a {\it first order}
transition~\cite{page}, with the confinement-deconfinement 
phase transition of the Boundary large $N_c$ $U(N_c)$ gauge theory 
at {\it finite temperatures}. The temporal Wilson lines, $P$, acquire 
non-zero vacuum expectation values above $T_0$, thereby inducing a spontaneous breaking of the center of the gauge group, 
while the {\it spatial} Wilson loops acquire an {\it area} law
below $T_0$,  
with (constant, temperature independent) 
string tension $T_{str}$, given by the square of the (large) 
AdS Radius of curvature:
\be
     T_{str} \propto b^2 \propto (-\Lambda)^{-1} 
\label{stringtension}
\ee
in units of an elementary string tension (e.g. type IIB string in the 
approach of \cite{witten}).

\subsection{World-sheet defects, AdS space-times, 
and application
to Quark Confinement in non-supersymmetric theories} 

Let us now examine the 
above phase diagram from a world-sheet defect point of view.
We first mention that 
the inclusion 
of world-sheet defects   
then, the above picture of obtaining a finite string tension, is valid also in {\it non supersymmetric} theories. 
Indeed,  
at the position $\sigma _0$ of the defect 
(center of the loop $C$),  
where the target-space field $X(\sigma )$ diverges
(\ref{defect},\ref{spike})
the quantity $\sigma ^M$ (\ref{fourcomp}), entering 
the definition of the field  $\Phi$ (\ref{composite}),  
acquires contributions also from the bosonic parts
$\partial _a X^M$, 
since in that case there is a non-zero vorticity:  
\be
\epsilon ^{ab}\partial_{a}\partial_{b}X^M \ne 0
\label{vorticity}
\ee
It is important to notice that condensation of 
composite operators containing (\ref{vorticity}),  
e.g. appropriate 
powers as in (\ref{composite}), 
can
occur for {\it both} vortex (\ref{solution}) 
and monopole (\ref{solution2}) configurations~\footnote{The quantity 
(\ref{vorticity}) is non trivial for vortices, since 
the vortex angular variable is not differentiable 
due to its 
non-trivial winding number around a closed loop. Also 
(\ref{vorticity}) 
is not 
well defined at the 
monopole core, and requires 
regularization, e.g. by cutting small loops around the singularity.}. 
This allows the passage to 
the non-critical-dimension string picture of $QCD$, sketched in section 
3.2.

We also note that in a dilute-gas approximation 
for the defects,
where only a single defect  is considered
on the world-sheet, the two-dimensional surface $\Sigma$
with boundary $C$ which encircles the defect does not contain 
any other defects.
In that case, the quantities $\sigma^M$ have 
a non-supersymmetric counterpart, due to the non-trivial 
vorticity (\ref{vorticity}).
However, in a situation of many defects, as we expect to be the case 
beyond dilute gas approximation, 
one encounters defect singularities on $\Sigma (C)$.
For such cases, a {\it non-supersymmetric
remnant} of the observable (\ref{second}) exists, and  
the pertinent 
world-sheet action (\ref{second}) 
equals the standard bosonic string Nambu-Goto action,
upon (\ref{tension}).

Let us 
first examine the case studied in \cite{witten}, 
which is based on critical-dimension (super)strings. 
In that case, one may consider the role of vortices on the world-sheet 
of the string, wrapped around the compact dimension 
$S^1$ in the manifold with topology $X_1$, where such configurations
exist~\cite{sathiap}. 
Then, 
condensation  of such vortices, bound into pairs with antivortices, 
results in the quantity $\Phi $  (\ref{composite})
acquiring a non-trivial v.e.v. $<\Phi > \ne 0$.
In the {\it classical} AdS picture, described above, 
it is the non-trivial Planckian dynamics 
of a five-dimensional space time which is responsible for the above 
phenomenon. In such a case the standard BKT phase transitions 
of {\it vortex} condensation 
on the world 
sheet of {\it critical} strings~\cite{sathiap},
may be identified with $T_0$ of the AdS Black Hole space time. 
If then this  transition 
is associated with the traditional `Hagedorn' temperature,
where vortex fields become tachyonic~\cite{sathiap}, 
then one has a nice (non-perturbative)
picture where the confining phase transition 
coincides with the Hagedorn temperature. 

In the picture advocated in section 2, where both types of defects are present,
one has a nice correspondence of the entire phase diagram of $AdS$ Black Holes
presented above, with the 
the higher temperature $T_2$ -associated with 
Planckian scales in the $AdS_5$ space time- 
corresponding to  
{\it monopole condensation}. 
This would define an effective Regge slope for the string
\be
   \alpha ' = \frac{8\pi^3}{m_p^{1/2}(-\Lambda)^{3/4}}
\label{regge}
\ee
which may be thought of as the inverse of a string tension. 
Then, the definition of the string tension   
(\ref{stringtension}) is consistent with (\ref{regge}),
provided one identifies 
$m_p \sim (-\Lambda)^{-7/2}$ in some fundamental string units.
Notice that, in this way, $m_p^2 >> |\Lambda| $,
(\ref{regge}) is small, and the entire approach is consistent. 
In this picture, then, one has a 
higher-dimensional analogue of the correspondence between 
world-sheet monopole/anti monopole pairs 
and Schwarzschild Black Holes in two-target-space dimensional
strings~\cite{emnmonop}.  

The above phase diagram applies intact 
also to
the non-critical string approach to Gauge Theory, advocated 
in section 3.2. However, in this more general 
case one may think of the scales 
$-\Lambda$ and $m_p$ as being independent. As we shall discuss later on,
in our Liouville approach,  
the quantity $m_p$ may also receive contributions from 
{\it microscopic} fluctuations of 
the $AdS$ space.
Hence, in this picture, one  
obtains the following phase diagram (c.f. (\ref{regions3}) 
for 
the non-Abelian gauge theory, or a compact Abelian Gauge theory:
\bea
&~&  (i)~   T < T_{BKT-vortex}:~~~{\rm Confining~Phase}\nn \\
&~& (ii)~ T_{BKT-vortex} 
< T < T_{BKT-monop}:~~~~{\rm Deconfining~phase} \nn \\&~&  
{\rm where}~~ T_{BKT-monop} >> T_{BKT-vortex} \nn \\
&~& (iii)~ T > T_{BKT-monop}~~~{\rm Planckian~Physics}
\label{Pphys}
\eea
Notice that in this approach,
the string tension appearing in 
the Planckian Physics regime, due to 
world-sheet monopole condensation (corresponding to stable Black Holes
in $AdS$~\cite{page}), is not the same 
as the confining string tension in the low-energy 
phase of the theory; the Planckian tension 
may be viewed as a fundamental string tension,
due to quantum-gravitational effects. 
The existence of a higher temperature, 
such as $T_2$ or the monopole condensation 
temperature in our
world-sheet picture, 
which sets the scale 
for new Physics (Planckian), 
and hence may lead to a unified picture
involving effective and fundamental (grand-unified) strings,   
is not in disagreement with generic 
conjectures made in some 
simple models of QCD, based on the 
effective Lagrangian approach~\cite{ellis,ellis2}. 
Also notice that 
in such models, the gluon transition had been identified with the 
deconfinement transition, at least for large number of colours, 
and argued to be of first order.

There still remains the issue on the nature of the 
$T_1$ intermediate
temperature, appearing in the phase diagram of \cite{page};
In view of the vortex-monopole duality 
(\ref{exchange}), as well as the  
discussion preceding (\ref{sd}),
one would 
be tempted to 
identify $T_1$ (\ref{t1}) with the 
temperature corresponding to the self-dual radius. 
This, however, does not work. 
Indeed, if such were the case then the following relation should be valid
(c.f. (\ref{connect})):
\be 
      T_2=T_1^2/T_0
\label{relation}
\ee
which, in view of (\ref{crit}),(\ref{t1}),(\ref{t2})
would imply 
$m_p = \frac{4}{9\pi^2}(-\Lambda )^{1/2}$.
This contradicts the assumption $m_p^2 >> -\Lambda $ on which 
the above analysis is based~\cite{page}. Notice, in this respect, 
that 
the effective Regge slope (\ref{regge}), combined with (\ref{stringtension}), 
defines a self-dual temperature
$T_{sd} \propto (-\Lambda )^{-1/2} $ such that $T_2 >> T_{sd} >> T_1 > T_0$
for $|\Lambda |$ small.

However, there may not be a phase transition at $T_1$. 
Indeed,  in the region $T_0 < T < T_1$ the black hole 
has a tendency to evaporate (but in a unitary way),
whilst above $T_1$ the radiation tunnels to the Black Hole space time.  
A possible scenario describing the pertinent tunneling phenomena 
would be that there exist some other relevant operators in the 
world-sheet theory, e.g. instantons, which cause this instability;
indeed that picture would match the picture in the two-dimensional 
black hole case~\cite{emn,yung}, with a cigar-type target-space 
metric to which a 
world-sheet monopole-antimonopole pair is equivalent~\cite{emnmonop}.
For such metrics, there are instanton solutions on the world-sheet,
which constitute relevant operators. This was argued in ref. \cite{emn} to 
describe transition among black hole states.  
Notice that the spherically symmetric part of the metric for the four-dimensional AdS pf \cite{page},  
(\ref{adsbh}),(\ref{V}) approaches a cigar type metric for $b >> 1$,
so one may conjecture that at least 
in the limit of large $b$ 
there are world-sheet instantons in the $n$-dimensional  
case too, in the corresponding $\sigma$-model. 
A formal construction of an exact conformal field theory 
for such Black Hole space times, which would be 
the higher-dimensional analogue of \cite{wittenbh}, 
is still eluding us, but we think 
that the present work offers 
a sufficient number of  arguments,
and non-trivial consistency checks, 
in favour
of its existence.

We now note that if we 
accept the above picture, and associate the string tension
with the $AdS_5$ radius, $b^2$, then 
in our `recoil'-induced anti-de-Sitter space 
 $b^2 \propto |\epsilon|^{-4}$. 
Notice that $|\epsilon |^{-2}$ is
proportional (\ref{epsilon}) to the size of the world-sheet disc~\cite{kmw}. 
If we interpret such world-sheet discs as the minimal area enclosed 
by Wilson loops, we may then consider 
the limit $\epsilon \rightarrow 0^+$, where the analysis of \cite{kmw}
can be trusted, 
as representing {\it macroscopic} loops, 
appropriate for quark confinement. 
If one, then, applies the approach of \cite{witten}
to this case, by identifying the string tension with the 
squared radius of the $ADS_5$, one obtains a 
logarithmic scaling violation of the area law, manifested as a 
dependence of the string tension on the 
logarithm of the area of the large quark loop:
\be
    T_{str} \sim {\rm ln}^2A, ~~~~~A={\rm Loop~minimal~area}
\label{loop}
\ee
Then, 
following (\ref{tension}), one would be tempted to interpret such a 
result 
as reflecting an {\it asymptotic freedom} 
on the coupling $\kappa _1$. As we mentioned previously, 
this coupling is expected to be proportional to the gauge coupling 
of the original theory. The fact that the confinement 
occurs at the region of strong gauge couplings 
constitutes a nice consistency check of the approach. 

We now mention an alternative
possibility, which 
may also be responsible for the 
appearance of a non-vanishing string tension
in the sense of (\ref{tension}).
Instead of large AdS radii, 
one may consider 
{\it microscopic} 
five-dimensional $AdS_5$ black holes, 
whose intersection with the surface $\Sigma (C)$ bounded by the macroscopic 
Wilson loop, would 
correspond to small 2-d regions bounded by microscopic loops
(of the size of the Planckian length, which is the effective
UV cut-off in the target-space 
low-energy theory). In our D-string picture, 
such microscopic loops correspond formally to $|\epsilon |^2 \rightarrow 
\infty$, 
viewed as a world-sheet renormalization 
group scale. 
Although in such a case the approximations of \cite{kmw} break down, 
however, one may consider a formal extension of the above arguments; 
in that case, such small loops on $\Sigma (C)$,  
could be viewed as `regularized' world-sheet defects. 
Our Liouville theory approach, then,  
applies, at least formally, to this case as well.
However, the thermodynamical analysis 
of \cite{page} may not be valid. 
If one, accepts though, 
the existence of a temperature $T_0'$, which induces topology changes
for the microscopic fluctuations of space time, then such a 
temperature could be identified
with the corresponding monopole 
condensation BKT transitions in our Liouville 
approach~\footnote{From the point of view of the gauge
theory~\cite{malda,witten}, where the $AdS$-space radius of curvature is 
proportional to some 
positive power of $g_{YM}^2 N_c =\alpha $, kept fixed 
as $N_c \rightarrow \infty$,  
the region of large $\epsilon$ corresponds 
to weak coupling, $\alpha << 1$, 
and, thus, to the  high-energy region, where the gauge theory 
is asymptotically free. In this regime, the standard BKT 
world-sheet 
phase diagram, with three temperatures, for monopole and vortex condensation,
and self-dual situation, 
satisfying (\ref{relation}), may be valid. However, the string background 
computations 
cannot be trusted.}. In this regime, the string tension 
could arise via (\ref{tension}),
through,say, condensation 
of these microscopic fluctuations (monopoles in a world-sheet picture).  
However, the string tension in such a case is expected to depend 
on the microscopic AdS radii $b^2 \lsim L_{Plancl}^2$, and thus 
it will be much weaker than the one 
obtained in 
the macroscopic $AdS_5$ case; 
this is why we expect this latter  
approach, if correct, 
to be relevant for the appearance of 
a string tension in grand-unified strings.  
Certainly, such non-trivial issues require further investigations
before any serious 
claims can be made.

\subsection{Abelian Projection, Abelian Dominance Hypothesis 
 and Quark Confinement}

As already mentioned, one expects the  
above 
connection of Wilson loops 
with `black hole 
(classical) physics' 
to be valid only for Non-Abelian gauge theories or compact Abelian 
Gauge theories with (magnetic) monopoles.
To show this formally in the non-Abelian case
would necessitate the extension of 
the formalism of \cite{awada} to take into account Path ordering in the 
definition of the Wilson Loop, or the second superstring observable,
which is technically  
a non trivial task. However we expect no conceptual problems
with such an extension.

At the present stage, we have  
a formal proof of the equivalence between 
gauge fields and strings, in the sense of (\ref{awadaabel}),  
only for 
Abelian gauge fields~\cite{awada}. 
The confinement mechanism, described above, 
is valid only for non-Abelian gauge fields, or 
compact Abelian gauge fields, with magnetic monopoles. 
For other cases, although the formal equivalence (\ref{second}) 
may be valid, however, the resulting string tension (\ref{tension}) 
should be zero. In our approach above this can be understood by the fact that
the resulting scale invariant string appearing in (\ref{second}) lives 
in a flat target space, in contrast to the non-Abelian case, where 
the bulk $AdS$ space, associated 
with it in the sense of \cite{malda,witten},  
has been conjectured to have non-trivial topologies. 

However, even at the present stage, 
one is able of making 
some remarks on the long-distance physics concerning quark confinement,
as we shall argue now.
This is due to  
`t Hooft's observation on 
the r\^ole of Abelian gauge theories 
on the low-energy confining physics. 

For definiteness we shall  
consider the case 
of the Non-Abelian gauge group 
$U(N_c)$, $N_c \rightarrow \infty$, gauge theories. 
One may invoke the `Abelian Projection' conjecture  of `t Hooft~\cite{thooft}, 
according to which one can fix the gauge in a non-Abelian 
gauge theory, in such a way that one can remove as many 
non-Abelian degrees of freedom as possible, in the sense that 
only a {\it maximal torus group H} of the gauge group G remains unbroken.
For the group G=$SU(N_c)$, for instance, $H=U(1)^{N_c-1}$.
According to `t Hooft' the Abelian projected theory 
reduces to a $U(1)^{N_c-1}$ abelian gauge theory {\it supplemented} by a 
{\it magnetic monopole}. In such a case, according to  
Mandelstam, 
confinement manifests itself as {\it superconductivity} 
(type II) in the dual theory, in the 
phase where  monopole condense.
In addition, to the Abelian projection hypothesis, 
a stronger statement had also been made in the literature,
the so-called Abelian dominance hypothesis, according to which  
only the Abelian parts of a non-Abelian gauge group  
play an important r\^ole for confinement of quarks~\cite{ezawa}. 
For a recent discussion on Abelian -projected QCD and confinement 
see \cite{kondo}.
An important ingredient in all these works is the
decoupling 
of the `non-Abelian' degrees of freedom at the expense of 
having 
magnetic 
monopoles in the dual theory. 

We now come to our stringy case, for the gauge group G=$SU(N_c)$,
for large $N_c$. The theory is conformal at zero temperatures,
and leaves in the boundary of a dynamically 
appearing $AdS_5$ space, as explained above. 
According to the previous discussion, 
one may fix the 
gauge to the 
Abelian projected one, in which only the subgroup 
$U(1)^{N_c-1}$ matters, but there is a dual magnetic monopole. 
Then, the relevant 
Wilson loop or the superstring-like observable (\ref{second}) 
factorize 
\be
     <\Psi _1 (\Sigma)>^{N_c-1} =e^{-(N_c-1) S_{1,int}^{(2)}}
\label{apro}
\ee
where the subscript $1$ indicates quantities in a $U(1)$ gauge theory,
and we have passed into a Euclidean formalism. 
The large $N_c >> 1$ limit in such a case may be interpreted as implying an
effective 
string tension $T_{st} \equiv N_c {\kappa}_1^2 <\Phi>_1 \rightarrow \infty $, 
or a Regge slope $\alpha ' \sim T_{st}^{-1} \rightarrow 0$, therefore the low-energy field theory (perturbation expansion in powers of $\alpha '$) may be trusted. This is analogous, but not identical, to the arguments why 
large $N$ is needed for the analysis of \cite{malda,witten}.  
The relation (\ref{apro}) implies, then, Wilson's criterion for confinement,
if one interprets the world-sheet superstring action $S_{1,int}$ 
as an area of the pertinent loop.  
Notice here that, as a result of the 
spontaneous breaking of scale invariance, in the sense of the condensation 
(\ref{tension}), a confining area law makes perfect sense. 
This is the result of 
world-sheet vortex condensation.
The relevant confinement-deconfinement 
phase transition  
has been argued, following \cite{witten}, 
to 
occur due to topology change  
(appearance of singular 
space-time structures (black holes)) in the Liouville $AdS$. 
Notice that the presence of world-sheet vortex defects,
associated with non-zero vorticity (\ref{vorticity}),  
can be
understood, through the pull 
back (\ref{pullback2}), as implying a 
corresponding non-trivial {\it dual} magnetic monopole  gauge field in target 
space, necessary for confinement in Abelian-projected theories~\cite{thooft}.  
The dual nature of the target-space defect is associated with the fact 
that the magnetic flux  through the surface surrounding the Wilson loop
is associated with the second term in the rhs of (\ref{pullback2}), 
which is 
related to the world-sheet monopole contributions, as mentioned already.
The world-sheet duality (\ref{exchange}), which maps a monopole onto 
a vortex, is thereby associated with a standard 
electric-magnetic duality in target space~\footnote{From this 
point of view, 
our approach in this work 
may also be considered 
relevant to related studies on the r\^ole of magnetic monopoles
on the confining aspects of compact Abelian 
gauge field theories, and, in particular, it 
may offer 
a way to describe 
the interaction 
between electric 
and magnetic (monopole) loops in such theories~\cite{kogut}.  
In this latter respect, we would like to 
mention the work of \cite{zeni}, according to which 
the presence of massless point-like sources 
in compact $U(1)$ 4d theories, leads naturally to two-dimensional 
shock wave configurations. Their magnetic counterparts (monopoles)  
have been argued to be responsible 
for a Coulomb-gas behaviour of the vacuum,
exhibiting  
BKT-like transitions, as we find here. However, at present,
it is not clear to us 
how to make a formal correspondence  of our results with 
such 
analyses. This issue deserves further investigation.}.

At present we consider the above
discussion  as conjectural, given that a 
mathematically rigorous  correspondence
of the 
condensate (\ref{tension}) with  AdS black holes, in the non-Abelian case, 
although plausible, 
has not yet become available, due to the 
aforementioned technical 
path-ordering difficulties 
associated with non-Abelian Wilson loops. However, the 
non-trivial consistency of the above plausibility arguments on the Abelian 
Projection points to the validity of this assumption.

\section{Conclusions} 

In this work~\cite{em} we have made a proposal for a non-critical 
Liouville string approach to the long-distance 
physics of four-dimensional 
(conformal) non-Abelian Gauge theories.
We have given 
many plausibility arguments in its favour, and performed
some non-trivial consistency checks. A crucial 
ingredient in the approach was the incorporation of 
world-sheet defects, monopoles and vortices.
Their presence was necessitated by 
the fact that, 
the 
`matter' central charge of the pertinent Liouville theory, 
was found to be in the `dangerous ' region $1< {\cal D} < 9$.
We have identified the relevant contributions 
of the world-sheet defects
to the Wilson loop of the 4-d Gauge Theory, 
and argued that they are 
described by 
the second superstring 
observable of \cite{awada},
which in this case has non-supersymmetric remnants.

The association of the world-sheet 
defects with D-particle boundary deformations
on the (open) world-sheet, bounded by the Wilson loop curve,
lead to non-trivial recoil effects, argued to describe 
the `distortion' of space time due to the motion 
of the heavy test particle that circulates along the Wilson loop.
Such recoil effects have resulted in a dynamical 
appearance of five-dimensional Anti-de-Sitter space times, as a result
of appropriate Liouville dressing. 

We have presented plausibility arguments, supported by 
non-trivial consistency checks, that the incorporation 
of non-Abelian Gauge fields leads, by 
the above-described Liouville dressing, 
to Black Hole $AdS_5$ space times.

The various Berezinski-Kosterlitz-Thouless phase transitions 
of a gas of world-sheet 
defects have been associated with the 
thermodynamics of the $AdS_5$ black holes, 
following the argumentation of 
Witten~\cite{witten}, based on the AdS/CFT correspondence
proposed in \cite{malda}. 
The resulting phase diagrams for the gauge theory 
have shown the existence of a deconfining phase,
which occurs at the Hagedorn temperature,
the latter being defined as the temperature at which world-sheet 
vortices condense~\cite{sathiap}. 
Our analysis has also revealed the existence 
of an interesting phase structure above this transition, which 
is in correspondence with the Thermodynamics of the $AdS$ Black Holes
discussed in \cite{page}.
There exists a temperature, much higher than the vortex condensation 
temperature, which sets the scale for Planckian Physics,
at which world-sheet monopoles condense, and thus 
stable black holes occur in $AdS_5$. 
Such a result is not in disagreement with generic expectations 
from corresponding studies in the context of 
the effective Lagrangian approach to confining 
$QCD$~\cite{ellis,ellis2}, predicting a much higher temperature
where new physics comes in. From the point of view 
of this work, this temperature has been associated 
with the Hagedorn temperature of grand-unified fundamental strings.

There are many things in our approach that should be checked, before 
the above claims can be 
considered rigorous. The first is the extension of \cite{awada} 
to the non-Abelian case, which is plagued with technical difficulties. 
The second is 
the fact that a proper treatment of non-critical (Liouville) 
boundary dynamics 
is still lacking. Due to world-sheet reparametrization invariance, 
the Liouville field is not simply a scalar field, as 
we have tacitly assumed in the presentation of the above arguments. 
This leads to complications in imposing boundary conditions~\cite{ambjorn}, 
and may alter some of the above results.  
What we hope, however, is that at least the major 
qualitative features of the above analysis, which are relevant for 
long-distance confining physics, will survive 
the proper treatment of boundary Liouville dynamics, 
especially in the dangerous regions for the 
matter central charges,
which is still eluding us. 

Before closing, we would also like to point out 
that our Liouville approach 
to four-dimensional gauge theories, which made use of 
non-trivial (logarithmic) boundary `D-particle recoil' deformations,
appears to offer a unified 
description of how the string 
tension may arise in various regimes of string theory, including
high-energy (Planckian) string tensions, as a 
result of microscopic fluctuations of $AdS$ space. The latter 
might be relevant
to searches for the elusive $M$-theory description of strings~\cite{mth}.  

\section*{Acknowledgements} 

This talk is based on original research 
done in collaboration with John Ellis~\cite{em}. 
We would like to thank C. Korthals-Altes, A. Momen and M. Teper 
for useful discussions. From discussions with I. Kogan we 
understand  
that he has 
also been pursuing~\cite{ikogan}, independently, 
similar ideas on the r\^ole of world-sheet defects
and their association with the $AdS$ Black-Hole Thermodynamics.
However, our approach is different in that it makes use of
boundary Liouville strings, which allows 
a dynamical appearance, in a world-sheet framework, of 
the five-dimensional $AdS$ space.

\end{document}